# Automated Lane Change via Adaptive Interactive MPC: Human-in-the-Loop Experiments

Viranjan Bhattacharyya<sup> </sup> and Ardalan Vahidi<sup> </sup>, *Fellow, IEEE*

*Abstract*— This article presents a new optimal control-based interactive motion planning algorithm for an autonomous vehicle interacting with a human-driven vehicle. The ego vehicle solves a joint optimization problem for its motion planning involving costs and coupled constraints of both vehicles and applies its own actions. The nonconvex feasible region and lane discipline are handled by introducing integer decision variables and the resulting optimization problem is a mixed-integer quadratic program (MIQP) which is implemented via model predictive control (MPC). Furthermore, the ego vehicle imputes the cost of human-driven neighboring vehicle (NV) using an inverse optimal control method based on Karush–Kuhn–Tucker (KKT) conditions and adapts the joint optimization cost accordingly. We call the algorithm adaptive interactive mixed-integer MPC (aiMPC). Its interaction with human subjects driving the NV in a mandatory lane change (MLC) scenario is tested in a developed software-and-human-in-the-loop simulator. Results show the effectiveness of the presented algorithm in terms of enhanced mobility of both vehicles compared to baseline methods.

## I. INTRODUCTION

INTERACTION with other road users is one of the core aspects of driving which humans execute naturally. For the foreseeable future, autonomous vehicles will drive alongside human-driven vehicles and therefore, making the onboard computer capable of interaction becomes vital. Imparting such a *sense* into the decision making and motion planning algorithms has been explored in some of the recent autonomous vehicle planning research [1], [2], [3]. It may be inferred that motion plans which are generated without considering the *mutual influence* in driving, result in overly conservative behavior of the ego vehicle. In interaction intensive scenarios like mandatory lane change (MLC) and merging, this may result in inability to execute the required maneuver [4].

Received 30 October 2023; revised 1 November 2023 and 26 April 2024; accepted 9 September 2024. This work was supported in part by Clemson University's Virtual Prototyping of Autonomy Enabled Ground Systems (VIPR-GS), U.S. Army Center of Excellence for Modeling and Simulation of Ground Vehicles with U.S. Army DEVCOM Ground Vehicle Systems Center (GVSC), under Agreement W56HZV-21-2-0001. Recommended by Associate Editor F. Borrelli. *(Corresponding author: Viranjan Bhattacharyya.)*

This work involved human subjects or animals in its research. Approval of all ethical and experimental procedures and protocols was granted by Clemson University's Institutional Review Board under Application No. IRB2023-0619.

The authors are with the Department of Mechanical Engineering, Clemson University, Clemson, SC 29634 USA (e-mail: vbhatta@clemson.edu; avahidi@clemson.edu).

Digital Object Identifier 10.1109/TCST.2024.3478028

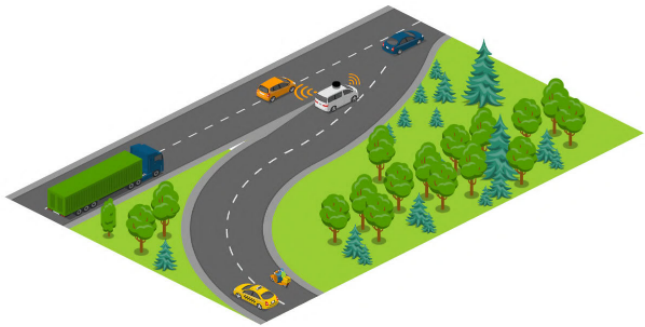

Fig. 1. Illustration of an interactive lane changing scenario.

In predictive motion planning, the most common approaches include constant velocity or acceleration assumption for the NV over the prediction horizon or assumption of intention sharing [5], [6]. These unilateral approaches lead to unnatural driving since in interaction intensive scenarios, these assumptions are unrealistic. Although the deviations from reality may be handled by modeling uncertainty [7], [8], [9], such uncertainties grow over the horizon and can lead to the "freezing robot problem." This has been recognized in [10] where a joint collision avoidance approach is presented. Based on [10], in this work, we aim to model the *mutual influence* by formulating a joint optimal control problem (OCP) solved by the ego vehicle. In particular, the joint problem is a mixed-integer quadratic program (MIQP) and defines *mutual influence* by delegating a control term of the neighbor in the OCP. Optimal trajectory planning via MIQP has been presented in previous research [5], [6], [11] but the mutual influence among vehicles has not been focused on. Duality-based smooth collision avoidance constraints have been developed in [12] but owing to nonconvexity, the solution depends on provision of an initial guess. MIQP can be solved with algorithms like branch-and-cut without the need for warm starting since MIQP formulation renders the subproblems in branch-and-cut as convex.

Previous research has explored modeling of the planning problem as a dynamic game [1], [2], [13], [14]. Some of the previous research works have also highlighted that if driving goals are modeled via cost functions, then the weights on various competing cost terms can capture the nature of driving. A model predictive control (MPC)-based human-like motion planning method with predefined cost weights to set the driving style is developed in [15]. The *nature* of interacting agents is estimated on a spectrum and a game adapted to it is solved online using MPC in [2]. However, these approaches involve nonconvex formulations. Furthermore, most of these works test the algorithms in simulation with prerecorded neighbor data. More recently, [16] presented a scenario tree-based interactive decision making Branch MPC which reduces driving conservativeness. The method was tested with modeled NVs in abstract simulations where real-time implementation aspects were not the focus. The method presented in [1] was tested in a relatively low-fidelity simulator with human subjects.

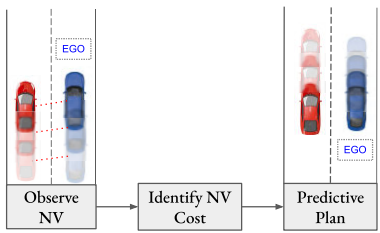

Fig. 2. Motion planning algorithm schematic.

In this work, we attempt to impart a sense of interaction into the motion planning algorithm of the ego vehicle (Fig. 1). Furthermore, our developed algorithm estimates the *nature* of the human-driven vehicle it is interacting with and acts accordingly. The idea of modeling human actions mathematically as OCP has been studied in works such as [17], [18], and [19]. The fundamental idea is to fit an OCP to observed trajectory data of human actions. This is the inverse OCP problem of finding the cost function given an "optimal" trajectory. Three common approaches [19], [20], [21] are compared in [22] in the context of inverse OCP. These methods are computationally expensive due to repeated solution of a forward OCP to identify the best cost parameters, hence not suitable for real-time inverse OCP. A computationally efficient method for continuous time models was presented in [22]. Taking inspiration from this, we develop a (convex) discrete-time online NV cost estimation method based on [23]. Although a method to combine the forward problem (without inequality constraints) and inverse problem has been developed in [24], we use MPC for solving the forward problem with explicit collision avoidance constraints and use convex residual minimization-based method [23] in the inverse problem for real-time implementability.

Fig. 2 illustrates the concept behind the algorithm: the ego vehicle observes the trajectory of a human-driven neighboring vehicle (NV), identifies its cost function (weights), and jointly predicts the future states capturing the *mutual influence*. The key contributions of this work (open sourced: https://github.com/autonomous-viranjan/Interactive-Motion-Planning) are

1) Modeling the interaction between an autonomous vehicle and a human-driven vehicle as a mixed logical dynamical [25] system and formulating the motion planning problem as a linear quadratic OCP.
2) Identifying a simple and explainable cost function of the human-driven vehicle and a method to adapt the problem online. The cost identification method is convex, hence, computationally efficient.
3) Testing with human subjects in a developed realistic 'in-the-loop' software simulation environment in real time.

The formulated interactive motion planning problem has linear dynamics and constraints, and quadratic cost, while the logical decision making is handled by integer variables. This makes our formulation amenable to solution with well-developed mixed-integer quadratic programming solvers [26].

This work addresses a few limitations of our previous work [3]. More specifically.

1) A core part of this work focuses on human subject experiments which did not exist in [3]. To conduct experiments with human subjects in real-time, the SHiL simulator was developed and is detailed in the article. Development of a high-fidelity (with respect to rendering, data communication, and modeling) simulator led to photorealistic visual feedback for the drivers which generated realistic interaction data. The code and data have been open sourced. This was a major task by itself with important results that were not reported in [3].
2) This work addresses a number of limitations of our paper's [3] formulation. We had attributed "aggressiveness" and "conservativeness" to velocity tracking and acceleration minimization cost, respectively. During human-in-the-loop experiments conducted for the current work, we recognized that this works only when the NV solves an MPC with the correctly assumed cost function. Human drivers do not solve an MPC and thus the formulation in our previous work did not lend itself well to actual human driving. To address this, the cost function in (8) has been redesigned for implementation. For NV cost estimation, a proximity and acceleration minimization-based cost has been designed. A detailed analysis of the cost imputation and resulting interactive motion with human driven NV is presented in Section V. The ego's new cost function prevents rapid acceleration and lane changes and enables auto-lane selection for the ego, which were important for implementation.
3) We extend the formulation to more than two lanes and more vehicles, which is another contribution over [3].

The article is organized as follows. In Section II, the formulation of the interactive motion planning problem as an OCP is discussed. In Section III, the NV cost estimation is presented. In Section IV, the test environment development and test design is described. In Section V, the results are discussed and analyzed. Finally, Section VI concludes the work and discusses the future challenges.

## II. PROBLEM FORMULATION

The interactive motion planning problem is formulated as an OCP of the ego vehicle solved as a joint optimization. The ego vehicle solves the OCP with the hypothesis that it has implicit *control* over the states of the NV and vice-versa, which captures the *mutual influence*, but executes only its own actions

$$\min_{u_{(\cdot)}, \beta^{(\cdot)}, \mu_{(\cdot)}} \sum_{i=k}^{k+N} \left[ J_{\text{ego}}(x_{\text{ego}}(i), u_{\text{ego}}(i)) \right.$$
$$\left. + J_{\text{NV}}(\hat{\alpha}, x_{\text{NV}}(i), u_{\text{NV}}(i), x_{\text{ego}}(i)) + q_{\text{slack}}\epsilon(i) \right]$$
(1a)

$$\text{s.t. } x_{\text{ego}}(i+1) = A_d x_{\text{ego}}(i) + B_d u_{\text{ego}}(i) \tag{1b}$$

$$x_{\text{NV}}(i+1) = A_{d,\text{NV}} x_{\text{NV}}(i) + B_{d,\text{NV}} u_{\text{NV}}(i) \tag{1c}$$

$$x_v(i) \in \chi_{\text{free}}\big(\beta^{(\cdot)}(i), \mu_1(i), \mu_2(i)\big) \tag{1d}$$

$$x_v(i) \in S_{\text{lane}}(\mu_1(i), \mu_2(i)) \tag{1e}$$

$$u_v(i) \in U \tag{1f}$$

$$\beta^{(\cdot)}(i), \quad \mu_1(i), \mu_2(i) \in \{0, 1\} \tag{1g}$$

where $J_v$ is the cost, $x_v$ are the states, and $u_v$ are the controls where $v \in \{\text{ego}, \text{NV}\}$ while $\beta^{(\cdot)}, \mu_1, \mu_2$ are binary logical variables. $\hat{\alpha}$ is the vector of weights, $\epsilon$ are the slack variables, $\chi_{\text{free}}$ is the safe region to avoid collision, $S_{\text{lane}}$ is the lane discipline set, and $U$ is the set of admissible controls. We introduce a method to estimate cost weights $\hat{\alpha}$ online through trajectory observations. The decision variables $u_{\text{ego}}$, $u_{\text{NV}}, \beta^{(\cdot)}, \mu_1$, and $\mu_2$ in ego's optimization model mutual influence while online estimation of $\hat{\alpha}$ facilitates estimation of the "type" of influence. We explain the decision variables in Sections II-A–II-C.

In this work, we specifically focus on longitudinal interaction between the ego vehicle and one human-driven NV in the neighboring lane. The motion planning problem solved by the ego vehicle is a MIQP implemented in a receding horizon manner (MPC).

This MPC is: 1) adaptive in nature as $\hat{\alpha}$ is estimated and adapted online; and 2) captures interaction with the joint cost, dynamics, and collision avoidance constraints which predict the trajectory of the ego and NV together—modeling the mutual influence over the MPC horizon. We use the abbreviation adaptive interactive mixed-integer MPC (aiMPC) to refer to this adaptive interactive MPC and its formulation builds on our previous work [3]. The actual NV state observation at every time step is fed to the MPC for reinitialization and acts as feedback to the controller.

In the following subsections, we describe the elements of the OCP (1) in detail.

### A. Vehicle Model

We use kinematic models for the vehicles since this is a high level planner. Furthermore, we use linear models for computational tractability and subproblem convexity.

*1) Ego:* We utilize the vehicle model developed by Dollar and Vahidi [5] which decouples the lateral and longitudinal dynamics

$$\frac{d}{dt}\begin{bmatrix} s \\ v_s \\ a \\ l \\ r_l \end{bmatrix} = \begin{bmatrix} 0 & 1 & 0 & 0 & 0 \\ 0 & 0 & 1 & 0 & 0 \\ 0 & 0 & -\frac{1}{\tau} & 0 & 0 \\ 0 & 0 & 0 & 0 & 1 \\ 0 & 0 & 0 & -\omega_n^2 & -2\zeta\omega_n \end{bmatrix}\begin{bmatrix} s \\ v_s \\ a \\ l \\ r_l \end{bmatrix} + \begin{bmatrix} 0 & 0 \\ 0 & 0 \\ \frac{1}{\tau} & 0 \\ 0 & 0 \\ 0 & K\omega_n^2 \end{bmatrix}\begin{bmatrix} u_a \\ u_l \end{bmatrix}. \tag{2}$$

TABLE I
PARAMETERS OF THE LONGITUDINAL DYNAMICS MODEL

| Parameter | Value |
|---|---|
| $\tau$ | 0.275 s |
| $\omega_n$ | 1.091 rad/s |
| $\zeta$ | 1 |
| $K$ | 1 |

The vehicle's longitudinal motion is modeled as a double integrator with the states position, velocity and acceleration, $[s, v_s, a]$, and time constant $\tau$, while the lateral motion is modeled as a second-order critically damped system with states lateral lane position and rate of change of lane, $[l, r_l]$. $\zeta$ is the damping ratio, $K$ is the gain, and $\omega_n$ is the natural frequency (Table I). The control inputs are longitudinal acceleration $u_a$ and lane command $u_l$ where $u_l$ is an integer equal to the commanded lane number. We discretize these dynamics when implementing in the controller as $x_{\text{ego}}(k+1) = A_d x_{\text{ego}}(k) + B_d u_{\text{ego}}(k)$ where $x_{\text{ego}} = [s\ v_s\ a\ l\ r_l]^T$ and $u_{\text{ego}} = [u_a\ u_l]^T$.

*2) NV:* The NV is modeled with the following assumption:

*Assumption 1:* The NV continues in its own lane and its motion can be modeled as

$$\frac{d}{dt}\begin{bmatrix} s_{\text{NV}} \\ v_{\text{NV}} \\ a_{\text{NV}} \end{bmatrix} = \begin{bmatrix} 0 & 1 & 0 \\ 0 & 0 & 1 \\ 0 & 0 & -\frac{1}{\tau} \end{bmatrix}\begin{bmatrix} s_{\text{NV}} \\ v_{\text{NV}} \\ a_{\text{NV}} \end{bmatrix} + \begin{bmatrix} 0 \\ 0 \\ \frac{1}{\tau} \end{bmatrix} u_{\text{NV}}. \tag{3}$$

These continuous time dynamics are discretized as $x_{\text{NV}}(k+1) = A_{d,\text{NV}} x_{\text{NV}}(k) + B_{d,\text{NV}} u_{\text{NV}}(k)$ where $x_{\text{NV}} = [s_{\text{NV}}\ v_{\text{NV}}\ a_{\text{NV}}]^T$ and $u_{\text{NV}}$ is the longitudinal acceleration control. Many urban roads and highways are two-lane in each direction and the NV would be restricted to its own lane when in the left lane faced with a mandatory lane changing autonomous vehicle on the right lane.

### B. Collision Avoidance Constraints

Due to the presence of obstacles, the drivable region $\chi_{\text{free}}$ is nonconvex. If in the same lane as an obstacle, the ego vehicle either needs to be ahead of the front of obstacle or behind the rear of obstacle. The Big $M$ method [27], in which $M$ is a large positive constant, is utilized to convert this "or" constraint to "and," making it amenable for mixed integer programming. In this work, we consider two kinds of obstacles—NV with which the ego interacts and a slow/stationary vehicle in the right-most lane.

For an NV in lane 2, over which the ego vehicle has mutual influence, the following joint collision avoidance constraints are enforced at each time step $k$:

$$s_{\text{ego}}(k) - s_{\text{NV}}(k) - M\beta^{\text{ego,NV}}(k) - M\mu_2(k) \geq d_{\text{gap}} - 2M \tag{4a}$$

$$s_{\text{NV}}(k) - s_{\text{ego}}(k) + M\beta^{\text{ego,NV}}(k) - M\mu_2(k) \geq d_{\text{gap}} - M \tag{4b}$$

$\beta^{\text{ego,NV}}$ is the ahead-behind indicator. The choice of $\beta^{\text{ego,NV}} = 1$ places the ego vehicle ahead of the NV in lane 2 while $\beta^{\text{ego,NV}} = 0$ places it behind. $\mu_2$ is the lane indicator. It is set to 1 if ego resides in the lane 2 and set to 0 if it does not. In effect, (4a) ensures that the ego vehicle maintains at least a $d_{\text{gap}}$ distance ahead of the NV when it is placed in the same lane and ahead of NV. When this is the case, (4b) gets inactive. On the other hand, when the ego vehicle is placed behind NV, (4b) ensures that it maintains at least the $d_{\text{gap}}$ distance behind it while (4a) is rendered inactive.

For a slow/stationary vehicle in lane 1 which is not interacting with the ego vehicle, we enforce the following collision avoidance constraints:

$$s_{\text{ego}}(k) - s_{\text{obs}}(k) + \epsilon(k) - M\beta^{\text{ego,obs}}(k) - M\mu_1(k) \geq d_{\text{gap}} - 2M \quad (5a)$$

$$s_{\text{obs}}(k) - s_{\text{ego}}(k) + \epsilon(k) + M\beta^{\text{ego,obs}}(k) - M\mu_1(k) \geq d_{\text{gap}} - M. \quad (5b)$$

The logic is same as constraint (4) with the addition of slack variable $\epsilon$. The slack variable allows the ego vehicle to translate closer to the obstacle obs at a distance $d_{\text{gap}} - \epsilon$. The slack variable is a decision variable and is minimized with large weight $q_{\text{slack}}$. This prevents the optimization problem from becoming infeasible in edge cases by "slightly" relaxing the gap constraints. We introduce slack variables only where necessary since adding decision variables results in increased computation time. Equations (4) and (5) together enforce (1d).

### C. Lane Discipline

The binary logical variables $\mu_L$ need to capture whether (1) or not (0) the ego vehicle is placed in lane $L$. The following constraints enforce this for each lane $L \in \{1, 2\}$ at every time step $k$

$$l_{\text{ego}}(k) + M\mu_1(k) \leq 2 - \delta + M \quad (6a)$$
$$l_{\text{ego}}(k) - M\mu_2(k) \leq 1 + \delta \quad (6b)$$
$$l_{\text{ego}}(k) + M\mu_1(k) \geq 2 - \delta \quad (6c)$$
$$l_{\text{ego}}(k) - M\mu_2(k) \geq 1 + \delta - M. \quad (6d)$$

Here, $\delta = 0.5$ is the deviation of ego vehicle from the lane center up to which it is considered to be in that lane.

When ego resides in lane 1, $l_{\text{ego}} \leq 1.5$ so (6a) and (6b) enforce $\mu_1$ to be equal to 1 and $\mu_2$ to be equal to 0 which makes (6c) and (6d) inactive. Conversely, when ego resides in lane 2, $l_{\text{ego}} \geq 1.5$, so (6c) and (6d) enforce $\mu_1$ to be equal to 0 and $\mu_2$ to be equal to 1 while (6a) and (6b) are rendered inactive.

The experiments in this work consider two lanes and this implements the lane disciple constraint (1e).

### D. Control Admissibility

The acceleration control admissibility constraints combine velocity with the acceleration command to prevent operation in mechanically infeasible regions as explained in [28]

$$u_a \geq u_{a,\min} \quad (7a)$$
$$u_a \leq m_1 v + b_1 \quad (7b)$$
$$u_a \leq m_2 v + b_2. \quad (7c)$$

Here, the maximal acceleration command is velocity dependent and (7) is a convex approximation to the feasible region with two intersecting straight lines of slope $m_1 = 0.285$ s$^{-1}$ and $m_2 = -0.1208$ s$^{-1}$ with $u_1$ intercepts $b_1 = 2$ m/s$^2$, $b_2 = 4.83$ m/s$^2$, respectively, in the $(u_a - v)$ space.

### E. Cost Function

We model the joint cost as sum of costs of the ego and NV where

$$J_{\text{ego}} = q_v(v_{\text{ego}} - v_{\text{ref}})^2 + q_a a_{\text{ego}}^2 + q_a u_a^2 + q_{\text{da}}(\Delta a_{\text{ego}})^2 + q_{\text{dl}}(\Delta l_{\text{ego}})^2 + q_{\text{dl}}(\Delta u_l)^2 \quad (8a)$$

$$J_{\text{NV}} = \hat{\alpha}_p(s_{\text{NV}} - s_{\text{ego}})^2 + \hat{\alpha}_a[a_{\text{NV}}^2 + u_{\text{NV}}^2(\Delta a_{\text{NV}})^2]. \quad (8b)$$

The cost (8a) lets the ego vehicle optimize for the desired lane while the $\Delta l_{\text{ego}}$ term prevents rapid lane changes for lateral ride comfort. The acceleration state and control terms prevent excessive acceleration while minimizing $\Delta a_{\text{ego}}$ emphasizes longitudinal driving comfort. The neighbor model cost structure (8b) is chosen to capture the *longitudinal proximity* between NV and ego, (implicitly, the relative longitudinal velocity) and the acceleration of NV and its change. The rationale behind this cost design is as follows. Since we model the *interaction* via acceleration command i.e. the ego vehicle assumes that it can influence the longitudinal movement of NV and vice-versa through accelerations, when the proximity is minimized more, the mutual influence is high, and acceleration and its change can vary more. On the other hand, if the NV model's acceleration (and change) is penalized more, the mutual influence is lower and the proximity measure carries lower weight. In this work, we focus on estimating the NV's behavior, given the model and behavior parameters of the ego.

### F. More Than Two Lanes and Vehicles

The presented MIQP formulation is the generalizable to more than two lanes and vehicles

$$\min_{u(\cdot), \beta^{(\cdot)}, \mu_{(\cdot)}} \sum_{i=k}^{k+N} \left\{ J_{v=0}(x_0(i), u_0(i)) + \sum_{v=1}^{V} J_v(\hat{\alpha}_v, x_v(i), u_v(i), x_0(i)) + \sum_{v=1}^{V} J_v(\hat{\alpha}_v, x_v(i), u_v(i), x_0(i)) + q_{\text{slack}}\epsilon(i) \right\} \quad (9a)$$

s.t. $x_v(i+1) = A_{d,v} x_v(i) + B_{d,v} u_v(i)$ (9b)
$x_v(i) \in \chi_{\text{free}}(\beta^{v, \neg v}(i), \mu_L^v(i))$ (9c)
$x_v(i) \in S_{\text{lane}}(\mu_L^v(i))$ (9d)
$u_v(i) \in U$ (9e)
$\beta^{v, \neg v}(i), \quad \mu_L^v(i) \in \{0, 1\}$ (9f)

where cost of each vehicle $v \in \{0, 1, \ldots, V\}$ (ego := 0 and $V$ is the total number of NVs) is summed and the dynamics are concatenated. The states are $x_v = [s_v \ v_v \ a_v \ l_v \ r_{l,v}]^T$ i.e., longitudinal position, velocity, acceleration, lane position, and lane change rate. The controls are $u_v = [u_{a,v} \ u_{l,v}]^T$ i.e., acceleration command and lane command. $\hat{\alpha}_v$ are the cost weight vectors, $\epsilon$ are the slack variables and the binary decision variables are the front-back indicators $\beta^{v,\neg v}$ (neighbors of vehicle $v$ are denoted by $\neg v$), and lane indicators $\mu_L$ (for each lane $L$).

The collision-free region $\chi_{\text{free}}$ is defined by collision avoidance constraints as

$$s_v(k) - s_{\neg v}(k) - M\beta^{v,\neg v}(k) - M\mu_L^v(k) - M\mu_L^{\neg v}(k)$$
$$\geq d_{\text{gap}} - 3M \quad (10a)$$
$$s_{\neg v}(k) - s_v(k) + M\beta^{v,\neg v}(k) - M\mu_L^v(k) - M\mu_L^{\neg v}(k)$$
$$\geq d_{\text{gap}} - 2M \quad (10b)$$

where pairwise front-back indicator $\beta^{v,\neg v}$ is introduced for each vehicle and its neighbors. $M$ is a large positive number. When $v$ is in the same lane, $L$ as $\neg v$ ($\mu_L^v = \mu_L^{\neg v} = 1$): $\beta^{v,\neg v} = 1$ places vehicle $v$ ahead of $\neg v$ by at least the $d_{\text{gap}}$ ((10a) gets activated) while $\beta^{v,\neg v} = 0$ places vehicle $v$ behind $\neg v$ by at least the $d_{\text{gap}}$ [(10b) gets activated].

The lane discipline $S_{\text{lane}}$ constraints are enforced for each vehicle $v$ and lane $L$ as

$$l_v(k) \leq (L + \delta) + M(1 - \mu_L^v(k)) \quad (11a)$$
$$l_v(k) + M(1 - \mu_L^v(k)) \geq (L - \delta) \quad (11b)$$
$$\Sigma \mu_L^v(k) = 1 \quad (11c)$$

where $\mu_L^v$ is the indicator for each lane $L$ for vehicle $v$ and (11c) ensures that it resides in one lane at a time. $U$ is the admissible control set.

Fig. 3 shows a simulation with $J_0 = q_v(v_0 - v_{\text{ref}})^2 + q_a(a_0)^2$, $J_{1,2} = w_v(v_{1,2} - v_{\text{ref}})^2 + w_a(a_{1,2})^2$. However, it is observed that with the introduction of lateral degree of freedom, the merging of ego causes lane change of the NVs, resulting in low longitudinal interaction. This is quantified by low acceleration changes during merging [Fig. 3(b)].

## III. NEIGHBOR COST ESTIMATION

Inspired by the idea of online estimation of NV nature as presented in [2], we seek to infer the driving characteristics of the NV online by identifying its cost (weights) through trajectory observation. The cost parameters are estimated as presented in this section. Our approach imputes cost weights based on the observed trajectory data [23].

### A. Approximate Optimality

We aim to fit an OCP of the general form (12) to the observed trajectory data

$$\min_{x,u} f(x, u) \quad (12a)$$
$$\text{s.t.} \quad h_n(x, u, c_h) = 0 \quad (12b)$$
$$g_p(x, u, c_g) \leq 0 \quad (12c)$$

where $x \in \mathbb{R}^n$ are the state data, $u \in \mathbb{R}^m$ are the controls, $c_h$ and $c_g$ are constant parameters, $f(x)$ is the cost function, $h_n(x, u)$ are the equality constraint functions, and $g_j$ are the inequality constraint functions.

*Definition 1:* We define approximate optimality by defining residuals. Assuming that the observed trajectory is *approximately optimal* with respect to problem (12), the residuals

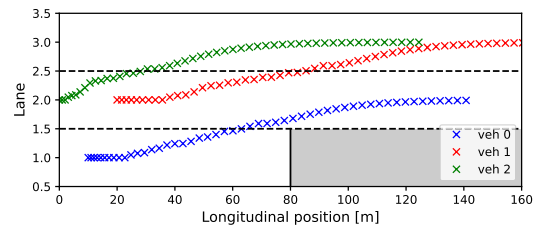

(a)

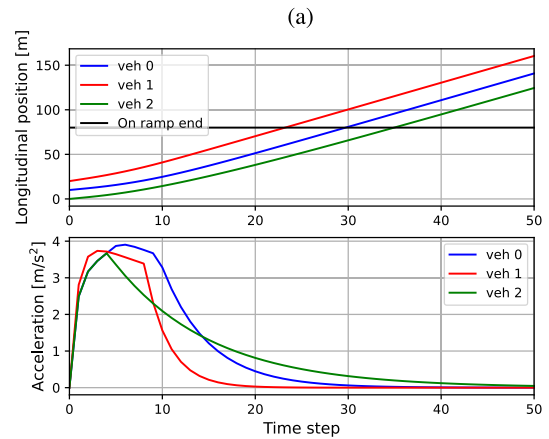

(b)

Fig. 3. Simulation with three lanes and three vehicles during merging: each vehicle is driven by a distributed MPC (with constant known cost) solving the problem in (9). (a) Vehicle trajectories. (b) Vehicle longitudinal positions and accelerations: low acceleration change during merging due to lateral degree of freedom.

based on the Karush–Kuhn–Tucker (KKT) conditions [29] are defined as

$$r_{\text{eq}} = h_n(x, u) \quad \forall n \quad (13a)$$
$$r_{\text{ineq}} = (g_p(x, u))_+ \quad \forall p \quad (13b)$$
$$r_{\text{stat}}(\alpha, \lambda, \nu) = \nabla f(x) + \Sigma_{j=1}^n \lambda_j \nabla g_j(x, u)$$
$$+ \Sigma_{j=1}^p \nu_j h_j(x, u) \quad (13c)$$
$$r_{\text{comp}}(\lambda) = \lambda_j g_j(x, u), \quad j = 1, \ldots, p. \quad (13d)$$

These residuals are deviations from the first-order necessary conditions of optimality.

Here, $r_{\text{eq}}$ and $r_{\text{ineq}}$ are the primal feasibility conditions which are independent of the dual variables. $r_{\text{eq}}$ are the deviation from equality constraints while $r_{\text{ineq}}$ are the deviations from the inequality constraints. The $+$ in (13b) denotes $g_p(\cdot)$ violating the inequality constraints by exceeding zero. $r_{\text{stat}}$ is the deviation from stationarity condition and $r_{\text{comp}}$ are the complementary slackness deviations.

*Assumption 2:* It is assumed that $r_{\text{eq}}$ and $r_{\text{ineq}}$ are close to zero i.e., the equality and inequality constraints are satisfied by the trajectory data and violations, if any, are negligible.

Now, for the observed trajectory to be *approximately optimal*, the residuals $r_{\text{stat}}$ and $r_{\text{comp}}$ are minimized. The dual variables $\lambda, \nu$ and the imputation parameter $\alpha$ are the decision variables and

$$f = \sum_j \alpha_j f_j, \quad \alpha = [\alpha_1, \ldots, \alpha_j] \quad (14a)$$

where $f_j$ are the basis functions for the cost function. Our goal is to find $\alpha$ for which $f$ is approximately consistent with the observed trajectory data.

### B. Estimation

We fit an OCP of the following form to the observed trajectory:

$$\min_{x_{NV}^{(i)}, u_{NV}^{(i-1)}} \sum_{i=k-r+1}^{k} \hat{J}_{NV}\left(\alpha, x_{NV}^{(i)}, x_{ego}^{(i)}\right) \quad (15a)$$

$$\text{s.t. } h_n\left(x_{NV}^{(i)}, x_{NV}^{(i-1)}, u_{NV}^{(i-1)}\right) = 0 \quad (15b)$$

$$g_p\left(x_{NV}^{(i)}, x_{ego}^{(i)}\right) \leq 0 \quad (15c)$$

where $k$ is the current time step and $r$ is the number of time steps in history for which the ego vehicle has observed the trajectory of the NV. $h_n$ are the NV vehicle dynamics and $g_p$ are the inequality constraints in the NV's (assumed) optimization problem. Other expressions have the same meaning as before and we denote trajectory data time step in the superscript. Essentially, we are fitting a problem of the general form (12) at each of the $r$ time steps.

With this, the residuals to be minimized are

$$r_{stat} = \Sigma_{i=k-r+1}^{k}(r)_{stat}\left(\alpha, \lambda^{(i)}, \nu^{(i)}\right)$$
$$= \Sigma_{i=k-r+1}^{k}\left[\nabla J_{NV} + \Sigma \lambda_p^{(i)} \nabla g_p + \Sigma \nu_n^{(i)} \nabla h_n\right] \quad (16a)$$

$$r_{comp} = \Sigma_{i=k-r+1}^{k} r_{comp}\left(\lambda^{(i)}\right) = \Sigma_{i=k-r+1}^{k} \lambda_p^{(i)} g_p. \quad (16b)$$

In our formulation, we fit a problem (15) such that only the state trajectory data of the NV and ego are required. Specifically, problem (15) is carefully chosen such that (16) results in the residuals containing only the state trajectory terms. This is important for practical implementation since the control data of NV may not be easily available. Here, $x_{NV}^{(k-r:k)}$ are the NV's observed trajectory data while $x_{ego}^{(k-r:k)}$ are that of the ego vehicle itself.

Finally, the following convex optimization problem is solved to impute the cost function based on the trajectory data:

$$\min_{\alpha, \lambda, \nu} \|r_{stat}\|_2^2 + \|r_{comp}\|_2^2 \quad (17a)$$

$$\text{s.t. } \alpha_p, \quad \alpha_a \geq 0 \quad (17b)$$

$$\lambda^{(i)} \geq 0 \quad (17c)$$

$$\alpha_p + \alpha_a = c \quad (17d)$$

where $c$ is a normalizing constant.

### C. Neighbor Behavior

We define $\hat{J}_{NV} = \alpha_p(s_{NV}^{(i)} - s_{ego}^{(i)})^2 + \alpha_a(a_{NV}^{(i)})^2$. With the imputed $\alpha = [\alpha_p \; \alpha_s]$, we set $\hat{\alpha}_p = \alpha_p$ and $\hat{\alpha}_a = \alpha_a$ in (8b). $\hat{J}_{NV}$ is designed with the rationale that high interaction arises at close proximity of ego and NV (high $\alpha_p$), and more influence is available due to low $\alpha_a$, allowing $a_{NV}$ to vary more. Conversely, low influence is available when $\alpha_a$ is high due to low variation in $a_{NV}$.

### D. Derivation of Cost Estimation

The terms in (15) are defined as

$$\hat{J}_{NV} = \alpha_p\left(s_{NV}^{(i)} - s_{ego}^{(i)}\right)^2 + \alpha_a\left(a_{NV}^{(i)}\right)^2 \quad (18a)$$

$$g_1 : 1 - \frac{\left(s_{NV}^{(i)} - s_{ego}^{(i)}\right)^2}{(L_d/2)^2} - \frac{\left(l_{NV}^{(i)} - l_{ego}^{(i)}\right)^2}{(W_d/2)^2} \leq 0 \quad (18b)$$

$$h_1 : s_{NV}^{(i)} - s_{NV}^{(i-1)} - v_{NV}^{(i)} \Delta t = 0 \quad (18c)$$

$$h_2 : v_{NV}^{(i)} - v_{NV}^{(i-1)} - a_{NV}^{(i)} \Delta t = 0 \quad (18d)$$

$$h_3 : a_{NV}^{(i)} - a_{NV}^{(i-1)} + \frac{1}{\tau}\left[a_{NV}^{(i-1)} - u_{NV}^{(i-1)}\right]\Delta t = 0 \quad (18e)$$

where $g_1$ is the collision avoidance constraint and $h_j$ are the assumed vehicle model for each state. $L_d$ is the vehicle length and $W_d$ is the width, and $\Delta t$ is the sampling time. The equality constraints $h_1, h_2$ are kinematic models and will be followed by definition while $h_3$ is a conventional first-order model. Therefore, $r_{eq}$ can be assumed to be close to zero. The ellipse in inequality constraint $g_1$ lies inside the vehicle edges and so, cannot be violated in reality since it requires overlapping of the vehicles. Hence, $r_{ineq}$ is assumed to be zero. This upholds assumption 2. We define the behavior estimating cost as in (18a): when the NV and ego are longitudinally very close, the *proximity* weight is higher; if acceleration of the NV is small, the *acceleration* weight is higher.

Hence,

$$r_{stat} = \sum_{i=k-r+1}^{k} \left[\nabla \hat{J}_{NV} + \lambda_1 \nabla g_1 + \nu_1 \nabla h_1 + \nu_2 \nabla h_2 + \nu_3 \nabla h_3\right] \quad (19a)$$

$$= \sum_{i=k-r+1}^{k} \begin{bmatrix} \frac{\partial \hat{J}_{NV}}{\partial s_{NV}} \\ \frac{\partial \hat{J}_{NV}}{\partial v_{NV}} \\ \frac{\partial \hat{J}_{NV}}{\partial a_{NV}} \\ \frac{\partial \hat{J}_{NV}}{\partial u_{NV}} \end{bmatrix}^{(i)} + \lambda_1^{(i)} \begin{bmatrix} \frac{\partial g_1}{\partial s_{NV}} \\ \frac{\partial g_1}{\partial v_{NV}} \\ \frac{\partial g_1}{\partial a_{NV}} \\ \frac{\partial g_1}{\partial u_{NV}} \end{bmatrix}^{(i)} + \nu_1^{(i)} \begin{bmatrix} \frac{\partial h_1}{\partial s_{NV}} \\ \frac{\partial h_1}{\partial v_{NV}} \\ \frac{\partial h_1}{\partial a_{NV}} \\ \frac{\partial h_1}{\partial u_{NV}} \end{bmatrix}^{(i)}$$

$$+ \nu_2^{(i)} \begin{bmatrix} \frac{\partial h_2}{\partial s_{NV}} \\ \frac{\partial h_2}{\partial v_{NV}} \\ \frac{\partial h_2}{\partial a_{NV}} \\ \frac{\partial h_2}{\partial u_{NV}} \end{bmatrix}^{(i)} + \nu_3^{(i)} \begin{bmatrix} \frac{\partial h_3}{\partial s_{NV}} \\ \frac{\partial h_3}{\partial v_{NV}} \\ \frac{\partial h_3}{\partial a_{NV}} \\ \frac{\partial h_3}{\partial u_{NV}} \end{bmatrix}^{(i)} \quad (19b)$$

$$= \sum_{i=k-r+1}^{k} \left\{ \begin{bmatrix} 2\alpha_p\left(s_{NV}^{(i)} - s_{ego}^{(i)}\right) \\ 0 \\ 2\alpha_a a_{NV}^{(i)} \\ 0 \end{bmatrix} + \lambda_1^{(i)} \begin{bmatrix} -2\frac{\left(s_{NV}^{(i)} - s_{ego}^{(i)}\right)}{(L_d/2)^2} \\ 0 \\ 0 \\ 0 \end{bmatrix} \right.$$

$$+ v_1^{(i)} \begin{bmatrix} 1 \\ 0 \\ 0 \\ 0 \end{bmatrix} + v_2^{(i)} \begin{bmatrix} 0 \\ 1 \\ 0 \\ 0 \end{bmatrix} + v_3^{(i)} \begin{bmatrix} 0 \\ 0 \\ 1 \\ -\Delta t/\tau \end{bmatrix} \Bigg\} \quad (19c)$$

and

$$r_{\text{comp}} = \sum_{i=k-r+1}^{k} \lambda_1 g_1 \quad (20a)$$

$$= \sum_{i=k-r+1}^{k} \lambda_1^{(i)} \left[ 1 - \left( \frac{s_{\text{NV}}^{(i)} - s_{\text{ego}}^{(i)}}{L_d/2} \right)^2 - \left( \frac{l_{\text{NV}}^{(i)} - l_{\text{ego}}^{(i)}}{W_d/2} \right)^2 \right]. \quad (20b)$$

Hence, the imputation problem is as follows:

$$\min_{\alpha, \lambda, v} \sum_{i=k-r+1}^{k} \left\| \begin{bmatrix} 2\alpha_p \left( s_{\text{NV}}^{(i)} - s_{\text{ego}}^{(i)} \right) - 2 \frac{\left( s_{\text{NV}}^{(i)} - s_{\text{ego}}^{(i)} \right)}{(L_d/2)^2} \lambda_1^{(i)} + v_1^{(i)} \\ v_2^{(i)} \\ 2\alpha_a a_{\text{NV}}^{(i)} + v_3^{(i)} \\ -\Delta t v_3^{(i)}/\tau \end{bmatrix} \right\|_2^2$$
$$+ \left\| \lambda_1^{(i)} \left[ 1 - \left( \frac{s_{\text{NV}}^{(i)} - s_{\text{ego}}^{(i)}}{L_d/2} \right)^2 - \left( \frac{l_{\text{NV}}^{(i)} - l_{\text{ego}}^{(i)}}{W_d/2} \right)^2 \right] \right\|_2^2 \quad (21a)$$

$$\text{s.t.} \quad \alpha_p, \quad \alpha_a \geq 0 \quad (21b)$$
$$\lambda_1^{(i)} \geq 0 \quad (21c)$$
$$\alpha_p + \alpha_a = c \quad (21d)$$

where the data time step is denoted in the superscript. Therefore, the imputation problem (21) is convex [although the forward problem is nonconvex due to (18b)].

## IV. TESTING WITH HUMANS

We test the proposed algorithm in a realistic software urban environment with human subjects driving a NV with which the ego vehicle interacts. We refer to this as a Software-and-Human-in-the-Loop (SHiL) simulator (Fig. 4).

### A. SHiL Development

The SHiL simulator is built on CARLA [30] and ROS [31]; the simulator is run on a Lenovo ThinkStation (Intel Xeon at 2.10 GHz + NVIDIA 32 GB Graphics) and the aiMPC is run on a Dell Precision (Intel Core i7 at 2.70 GHz), both with the Ubuntu 20.04.6 LTS operating system. CARLA is an open-source simulator for autonomous driving research with a client-server architecture and ROS is an open-source middleware which we use to set up communications between various software entities in the simulator. Fig. 5 shows the architecture. The CARLA Python API is used for defining and controlling various *actors* in CARLA. We design the architecture such that each major entity in the simulation is a ROS Node. Specifically, they are: Ego Vehicle Node, Planner Node, NV Node, and Obstacle Node.

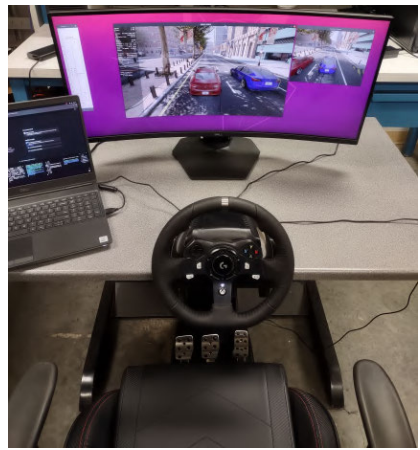

Fig. 4. SHiL setup.

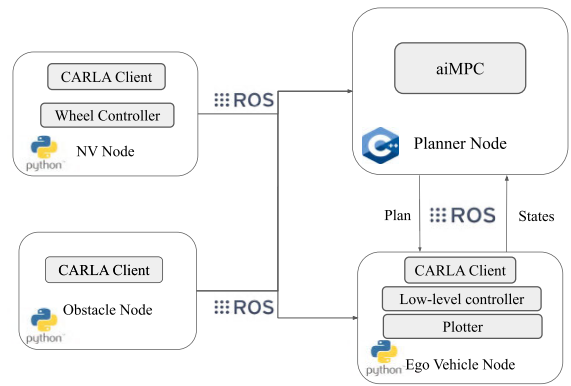

Fig. 5. SHiL architecture.

The architecture is explained below.
1) Planner Node and the MPC are implemented in C++ for real-time computation. The planner receives ego current states, NV states, obstacle states, and sends a planned trajectory via ROS.
2) The Ego Vehicle Node is a Python script which includes the following components: CARLA client, low-level controller, and a plotter. The low-level controller tracks the planner trajectory via PID control. The planned trajectory is received and ego states are fed back via ROS.
3) The NV Node includes a CARLA client and steering-pedal interface. It shares its current position and velocity via ROS.
4) The Obstacle Node includes a CARLA client and sends the position of the obstacle over ROS.

The aiMPC optimization is implemented using the GUROBI solver in C++. The NV is driven by human subjects via a Logitech G920 steering and pedal setup. pyLinuxWheel [32] and jstest-gtk [33] are incorporated to configure the pedals with the operating system. Pygame is used to receive commands from the steering-pedals and pass on to CARLA.

### B. SHiL Study

We test the developed adaptive interactive MPC planner algorithm in a MLC scenario (Fig. 6) on an urban road in the simulator. A stopped truck on the right lane necessitates lane change for the ego vehicle (blue). The human-driven NV (red) is on the left lane and the ego vehicle needs to negotiate with it to carry out the maneuver. Three subscenarios are created by changing the longitudinal position of the stopped truck. As discussed in literature [34], the most challenging interactive scenarios arise when the interacting vehicles start from similar positions and which one should yield is ambiguous.

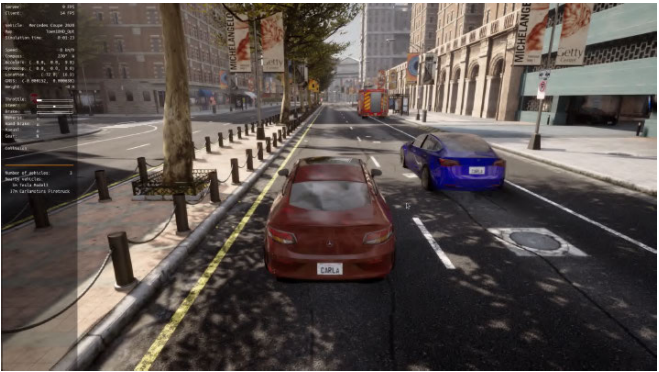

Fig. 6. Scenario illustration.

TABLE II
EXPERIMENT PARAMETERS

| Parameter | Value |
| --- | --- |
| Ego weights $q_v, q_a, q_{da}, q_{dl}$ | 10, 30, 100, 1000 |
| MPC horizon length $N$ | 20 |
| MPC sampling time $\Delta t$ | 0.2 s |
| Reference velocity $v_{\text{ref}}$ | 10 m/s |
| CARLA simulation time-step | 0.05 s |
| Imputation $r$ | 6 |
| Imputation interval | 6 iterations |

TABLE III
AVERAGE VELOCITY [m/s] COMPARISON

| Sub-scenario | | Baseline | Joint MPC | aiMPC |
| --- | --- | --- | --- | --- |
| 1 | $v_{\text{ego,avg}} =$ | 1.919 | 1.872 | 6.885 |
| | $v_{\text{NV,avg}} =$ | 4.332 | 5.346 | 7.612 |
| 2 | $v_{\text{ego,avg}} =$ | 2.384 | 2.040 | 7.210 |
| | $v_{\text{NV,avg}} =$ | 4.766 | 4.264 | 6.051 |
| 3 | $v_{\text{ego,avg}} =$ | 1.619 | 2.283 | 5.501 |
| | $v_{\text{NV,avg}} =$ | 3.905 | 4.857 | 5.048 |

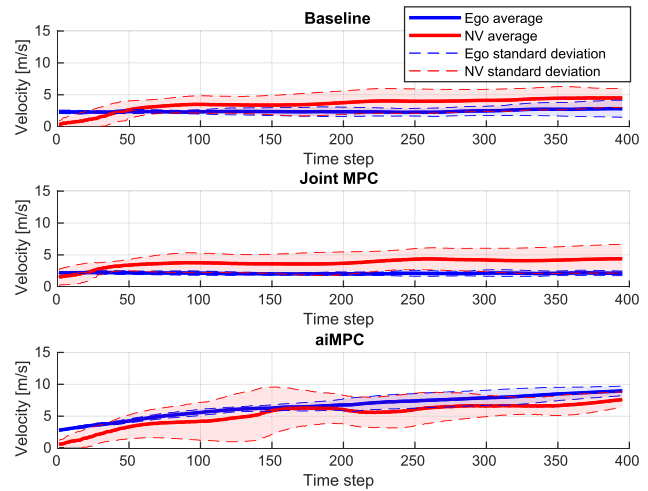

Fig. 7. Comparison of velocity trend of ego (blue) and NV (red) with the three algorithms.

The performance of the aiMPC is compared with.
1) A baseline MIQP MPC similar to [5] with a constant velocity model for NV's over the horizon trajectory prediction.
2) A joint MIQP MPC with the same formulation as in (1) but with equiweighted NV cost weights ($\alpha$) i.e., without online cost adaption.

The baseline constant velocity model, which is a common assumption, does not consider interaction at all. The second planner algorithm does consider the mutual influence via joint cost and control, but does not adapt itself based on NV's driving.

We recruited six volunteers to drive alongside the automated car (blue) in all three subscenarios. The human drivers were informed about the 10 m/s (36 km/h) speed limit and were asked to "drive as they would" in such scenarios. All SHiL experiments are run with the parameters (Table II):

In aiMPC, the ego observes NV for $r = 6$ steps and imputes its cost every 6 simulation steps.

## V. RESULTS AND ANALYSIS

Average velocities of the vehicles with all subjects in all subscenarios over the entire simulation are compared in Table III. The velocities of vehicles averaged over the subjects in subscenario 2 are shown in Fig. 7. Overall, the velocities of the ego vehicle are low in the baseline and nonadaptive joint MPC cases, which results in conservative motion. The ego vehicle velocities reach much closer to the reference of 10 m/s with the aiMPC planner which reduces the conservativeness. Intuitively, the aiMPC takes actions to "make space" for changing lane with the minimum compromise on velocity rather than just reacting to NV. Interestingly, we observe that the human-driven NV also moves faster with the aiMPC compared to driving with the other two planning algorithms. Therefore, the aiMPC reduces the conservativeness of the ego's motion and at the same time, influences the NV to reach velocities closer to the reference. This can be attributed to what is known as *social influence* in cognitive neuroscience [35].

With aiMPC, ego vehicle is able to deterministically adapt the MPC to changing lane ahead or behind by dedicating more weight to either *proximity* or *acceleration* for over-the-horizon joint predictions. Due to the NV trajectory-based estimation, the mode is determined by how the NV drives while how the NV drives is determined by ego's motion. Hence, rather than reactive actions, aiMPC is able to influence NV and adapt to its response.

### A. aiMPC Imputation

The imputation results for estimating NV's cost and the corresponding effect on ego's maneuver are shown in Fig. 8. In all three subscenarios, in line with our hypothesis, the aiMPC takes actions to adapt itself according to the NV and to influence it. The adaption results in the ego changing lane ahead of the subjects who yielded and behind the ones who did not. We observe that the aiMPC decides to change lane ahead or behind the NV based on the imputed cost weights $\alpha_p$ and $\alpha_a$. When the *proximity's* imputed weight $\alpha_p$ is high at the time of lane change, the ego vehicle overtakes the NV and changes lane ahead of it when it yields. Whereas, when the *acceleration's* imputed weight $\alpha_a$ is high during lane change maneuver, the ego changes lane behind NV.

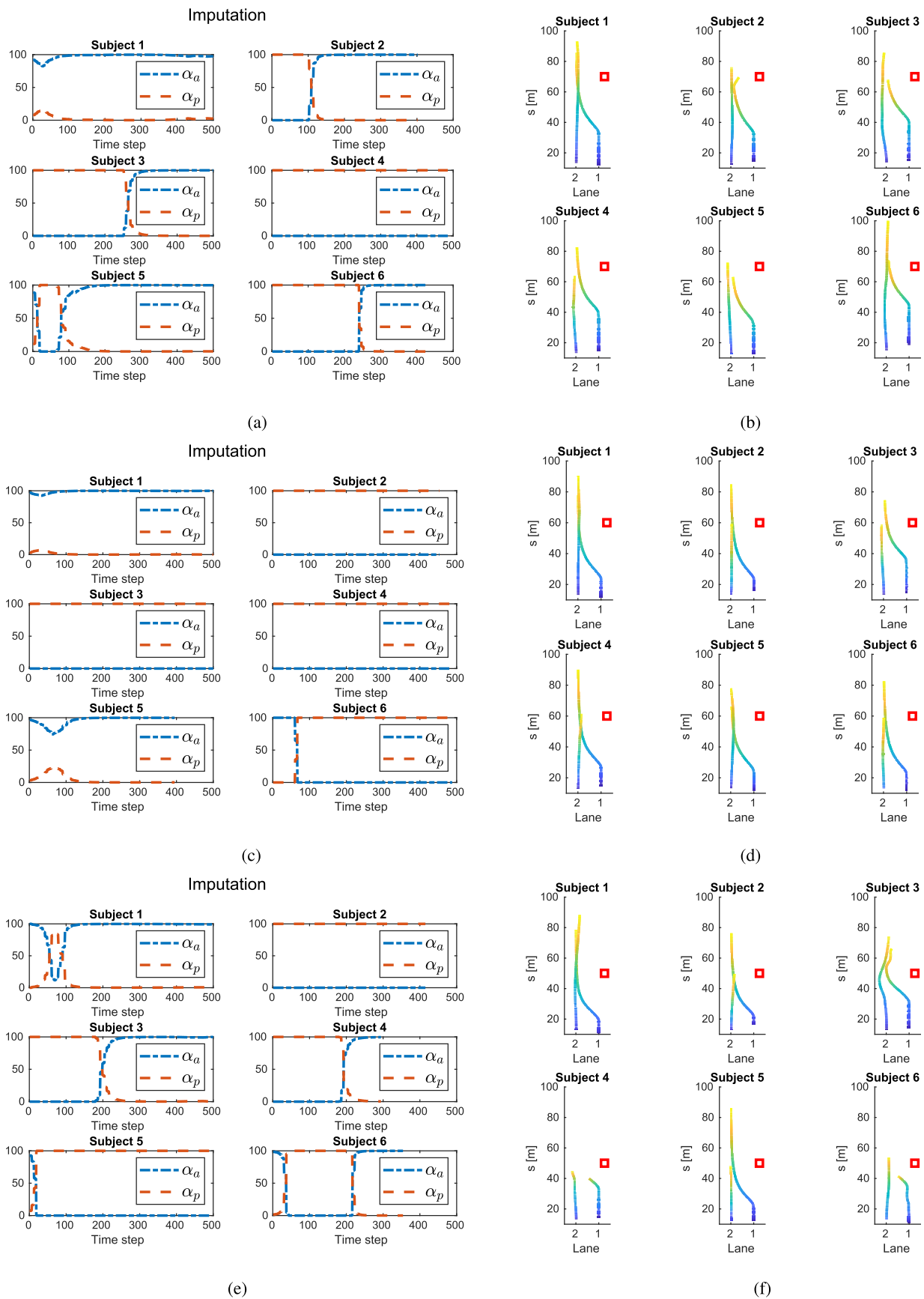

Fig. 8. MLC ahead or behind based on imputed $\alpha$. The imputed $\alpha_a$ and $\alpha_p$ are shown on the left and the corresponding ego and NV trajectory on the right. Ego starts in lane 1, NV starts in lane 2, and the trajectory becomes lighter as time progresses. Ego merges ahead when $\alpha_p$ of NV is high and $\alpha_a$ is low, and behind in the opposite case. (a) Subscenario 1: NV cost estimation. (b) Timed ego and NV trajectory. Obstacle at 70 m. (c) Subscenario 2: NV cost estimation. (d) Timed ego and NV trajectory. Obstacle at 60 m. (e) Subscenario 3: NV cost estimation. (f) Timed ego and NV trajectory. Obstacle at 50 m.

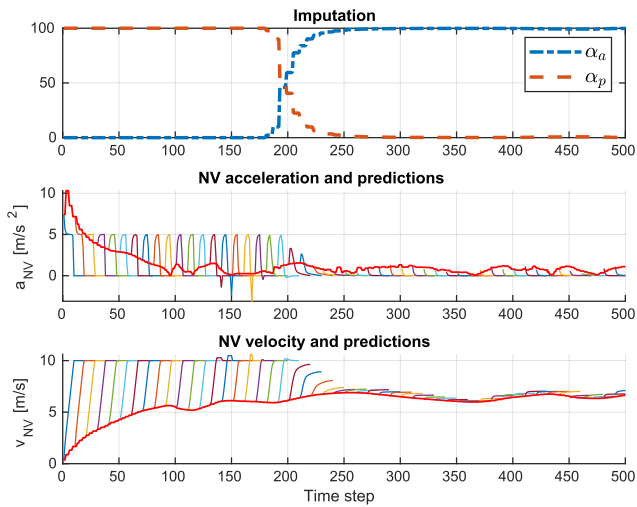

Fig. 9. Comparison of aiMPC predictions with change in $\alpha$: over-the-horizon acceleration and velocity predictions of the NV are shown emanating from actual values over time. High influence ($\alpha_p$) leads to larger predicted acceleration variation, while low influence tends to constant velocity predictions.

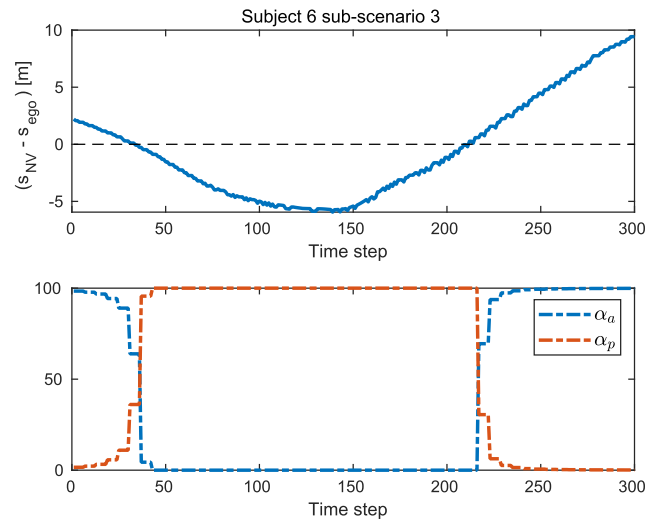

Fig. 10. $s_{\mathrm{NV}} - s_{\mathrm{ego}} = 0$ is a critical point for imputation of $\alpha$.

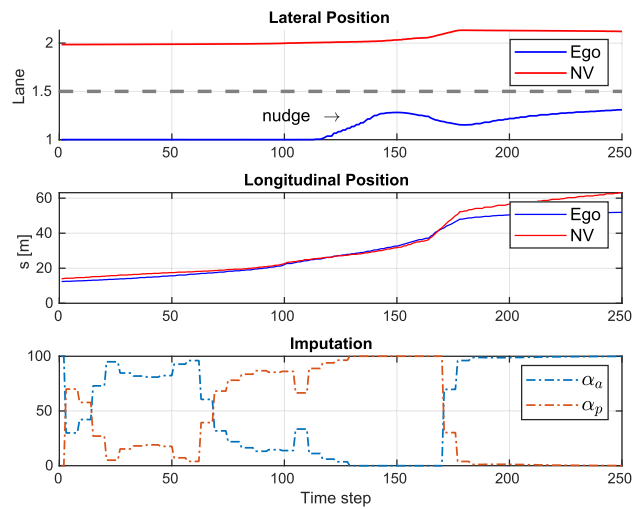

Fig. 11. Nudging example: Slightly before time step 150, $\alpha_p$ was imputed high so ego accelerated ahead and attempted lane change resulting in a nudge. However, NV did not yield and the constraints pulled the ego back toward lane 1. Slightly after step 150, due to NV's pacing ahead, $\alpha_p$ decreased and ego went behind NV.

In the residual minimization-based cost estimation in (21), the observed trajectory data determines the coefficients of the optimization variables. The imputation results can be described based on the *proximity* term $(s_{\mathrm{NV}} - s_{\mathrm{ego}})$ since its variation changes the relative weight (coefficients) of $\alpha_p$ and $\alpha_a$ in (21).

During imputation observation, if the NV yields and maintains close proximity, $(s_{\mathrm{NV}} - s_{\mathrm{ego}})$ is smaller so the coefficient of $\alpha_p$ in (21) is lower and imputed $\alpha_p$ is higher. Consequently, in the forward MPC problem (1) more weight is placed on longitudinal proximity minimization of the joint cost function (8) and the NV predicted acceleration can vary more due to lower $\alpha_a$, i.e., high *influence* is available. In these cases, the ego accelerates and maintains close longitudinal proximity over the horizon until the NV's acceleration drops enough to make space for it to change lane ahead. In the opposite case, when the NV paces ahead, $(s_{\mathrm{NV}} - s_{\mathrm{ego}})$ increases; the imputed $\alpha_p$ is smaller due to its larger coefficient. In the forward MPC problem, due to larger $\alpha_a$ more weight is placed on acceleration minimization of NV in (8) and as a result, low *influence* is available—tending to constant velocity over-the-horizon predictions of NV in the aiMPC.

This leads to solutions where the ego vehicle slows down and changes lane behind NV.[1] The acceleration and velocity predictions over-the-horizon for a subject from testing are shown in Fig. 9.

Switching between $\alpha_p$ and $\alpha_a$ occurs when there is a sign change in $(s_{\mathrm{NV}} - s_{\mathrm{ego}})$, i.e., one vehicle crosses the other longitudinally. This is illustrated in Fig. 10 with an example of one of the subjects from testing who switched back and forth. $(s_{\mathrm{NV}} - s_{\mathrm{ego}}) = 0$ is a critical point since at this point, the coefficient of $\alpha_p$ in (21) becomes zero.

### B. Nudging Behavior

The aiMPC is also able to generate nudges to attempt changing lane ahead of NV. When the NV maintains close proximity, $\alpha_p$ increases and ego attempts to change lane ahead. While the ego is attempting a lane change, if the NV driver decides not to yield and pace ahead, the ego is moved back to lane 1 due to the collision avoidance constraints. Such a situation results in a nudge and an example is plotted in Fig. 11.

### C. Computation Time

Over ten test trials (five cases each of ego lane changing ahead and behind), the average maximum computation time was 179 ms (201 ms absolute maximum) while the average computation time was 33 ms, on an Intel Xeon at 2.10 GHz processor. This lies within the aiMPC sampling time of 200 ms. In rare cases when the solution time exceeds 200 ms, the solver is set to return the solution obtained at the time limit of 200 ms.

---

[1]Sample trajectory prediction videos for both the cases are available at https://github.com/autonomous-viranjan/Interactive-Motion-Planning

## VI. CONCLUSION

This work presented a new interactive motion planning algorithm, aiMPC. The aiMPC solves an MIQP whose subproblems are convex, real-time implementable and enables the ego vehicle to take actions to influence the neighbor longitudinally and adapt itself based on it. We developed a realistic SHiL simulator to test this algorithm's interaction with a human-driven vehicle in the simulator in real-time. A two-lane MLC scenario was considered due to the high interactions involved. The presented method showed significant enhancement in mobility of both vehicles. The imputation method presented in this article is limited to longitudinal influence estimation. Future work could involve incorporation of lateral interaction, implementation, and testing of the algorithm on hardware for vehicle-in-the-loop experiments.

## APPENDIX A
## EGO VEHICLE LOW-LEVEL CONTROLLER IMPLEMENTATION

The vehicle states are converted from the CARLA frame to the planner frame using a coordinate transformation of the form: $X_{\text{HL}} = f_{\text{sim2hl}}(X_{\text{CARLA}})$, where $X_{\text{HL}} = [s\ v\ a\ l\ r_l]^T$ are the states in high-level (HL) frame with the same representations as in Section II and $X_{\text{CARLA}} = [x\ y\ v_x\ v_y\ a_x\ a_y\ \theta]^T$ where $(x, y)$ are the position coordinates, $(v_x, v_y)$ are velocities, $(a_x, a_y)$ are accelerations, in the respective directions, and $\theta$ is the yaw.

The high-level solution states and controls are passed to decoupled longitudinal and lateral PID controllers for tracking the acceleration command and lateral position reference, respectively. The PID controllers output the acceleration command $u_a$ and steering command $u_{\text{steer}}$, respectively. The acceleration actuation delay is modeled as

$$u_{\text{throttle}}(t) = u_{\text{throttle}}(t-1) \\ + \left[\frac{-1}{\tau_t}(u)_{\text{throttle}}(t-1) \\ + \frac{-1}{\tau_t}\left[u_a(t) + \frac{(u_a(t) - u_a(t-1))}{\Delta t}\right]\right]\Delta t \quad (22)$$

where $\tau_t$ is the delay. We present this directly in discrete time as this is an empirical relation.

## APPENDIX B
## HUMAN-DRIVEN NV

The following controller dynamics are augmented for the NV in addition to the CARLA internal vehicle dynamics (Fig. 12).

$$\dot{o}_{\text{steer}} = \frac{-1}{\tau_s}l + \frac{1}{\tau_s}(u_{\text{steer}} + \dot{u}_{\text{steer}}) \quad (23a)$$

$$\dot{o}_{\text{throttle}} = \frac{-1}{\tau_a}a + \frac{1}{\tau_a}(u_{\text{throttle}}) \quad (23b)$$

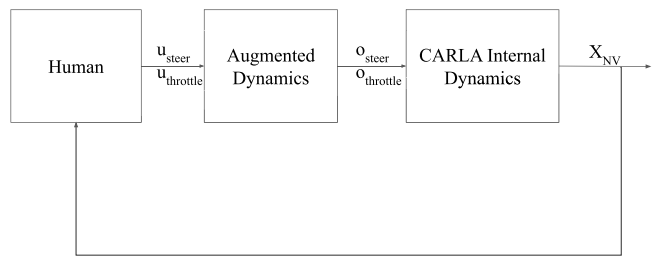

Fig. 12. NV control system: human is the controller, (23) are the dynamics augmented to CARLA internal vehicle dynamics.

where $o_{\text{steer}}$, $o_{\text{throttle}}$ are the NV steering and throttle actuation, respectively, $\tau_s$ and $\tau_a$ are the corresponding delays, $u_{\text{steer}}$ is the steering command from the G920 wheels, and $u_{\text{throttle}}$ is the throttle pedal command. Furthermore, for the lateral response, we add a derivative of steering command term to make the response faster based on rate of change of command. This adds a negative zero to the lateral dynamics of the system and creates the damping required for more realistic steering feel.

## ACKNOWLEDGMENT

DISTRIBUTION STATEMENT A. Approved for public release; distribution is unlimited. (OPSEC 8111). The authors would like to thank all volunteers. They would also like to thank Sydney Olson for helping with volunteer recruitment, Tyler Ard for helping with data analysis, and Dr. Austin Dollar for feedback on the formulation.

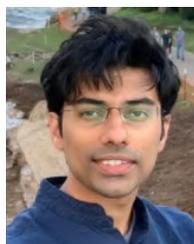

**Viranjan Bhattacharyya** received the M.S. degree in mechanical and aerospace engineering from Illinois Institute of Technology, Chicago, IL, USA, in 2021. He is currently pursuing the Ph.D. degree with the Department of Mechanical Engineering, Clemson University, Clemson, SC, USA.

His research focuses on model-based predictive motion planning and control of automated vehicles.

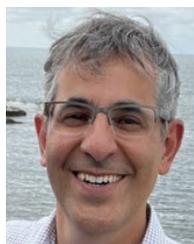

**Ardalan Vahidi** (Fellow, IEEE) received the B.S. and M.Sc. degrees in civil engineering from Sharif University, Tehran, Iran, in 1996 and 1998, respectively, the M.Sc. degree in transportation safety from George Washington University, Washington, DC, USA, in 2002, and the Ph.D. degree in mechanical engineering from the University of Michigan, Ann Arbor, MI, USA, in 2005.

He is a Professor of Mechanical Engineering at Clemson University, Clemson, SC, USA. His expertise is in optimal control applied to automated and connected vehicles, energy systems, and human performance. He has held scientific visiting positions at the University of California at Berkeley, Berkeley, CA, USA, from 2012 to 2013, BMW Technology Office, Mountain View, CA, USA, from 2012 to 2013, IFP Energies Nouvelles, Rueil-Malmaison, in 2017, and the University of California at San Diego, San Diego, CA, USA, in 2022.

Dr. Vahidi is a fellow of ASME.